\documentstyle[prc,aps,floats,epsfig,multicol]{revtex}
\tighten
\begin{document}
\draft

\title{Phi meson production in relativistic heavy ion collisions}
\author{Subrata Pal, C.M. Ko, and Zi-wei Lin}
\address{Cyclotron Institute and Physics Department,
Texas A\&M University, College Station, Texas 77843-3366}

\maketitle

\begin{abstract}
Within a multiphase transport model we study phi meson production 
in relativistic heavy ion collisions from both superposition of
initial multiple proton-proton interactions and the secondary collisions 
in the produced hadronic matter. The yield of phi mesons is then 
reconstructed from their decaying product of either the kaon-antikaon 
pairs or the dimuon pairs.  Since the kaon-antikaon pairs at midrapidity 
with low transverse momenta are predominantly rescattered or absorbed 
in the hadronic medium, they can not be used to reconstruct
the phi meson and lead thus to a smaller reconstructed phi meson 
yield than that reconstructed from the dimuon channel. 
With in-medium mass modifications of kaons and $\phi$ mesons, 
the $\phi$ yield from dimuons is further enhanced compared to 
that from the kaon-antikaon pairs. The model result is compared
with the experimental data at the CERN/SPS and RHIC energies 
and its implications to quark-gluon plasma formation are discussed.
\medskip

\noindent PACS numbers: 24.10.Cn, 24.10.Pa, 25.75.Dw
\end{abstract}


\section{introduction}

One of the major goals in relativistic heavy ion collisions is to study
the properties of hot and dense matter and possibly to create and identify
a new form of matter, the quark gluon plasma (QGP). It was suggested 
\cite{rafelski} that enhanced strangeness production could serve as 
an important signal for the deconfined matter. The dominant production 
of $s\bar s$ pairs via gluon-gluon interaction may lead to strangeness 
(chemical and flavor) equilibration times comparable to the lifetime 
of the plasma and much shorter than that of a thermally equilibrated 
hadronic fireball. The subsequent hadronization is then expected to 
result in an enhanced production of strange and multistrange particles 
and antiparticles. In particular, it has been argued that with
the formation of quark-gluon plasma not only the production of
phi mesons, which consist of $s\bar s$, is enhanced but they also 
retain the information on the conditions of the hot plasma as it is 
believed that phi mesons interact weakly in the hadronic matter 
and therefore freeze out quite early from the system \cite{shor}.

Using a proton or sulfur projectile at 200 GeV/nucleon and a
tungsten or uranium target, phi meson production has been previously 
studied at CERN/SPS by the NA38 Collaboration \cite{na38} and the HELIOS-3
Collaboration \cite{cern} via the dimuon invariant mass spectra.
The double ratio 
$( \phi /(\omega+\rho^{0} ))_{SU(W)} /( \phi / (\omega+\rho^{0}))_{pW}$ 
has been measured and was found to have a value of $2\sim 3$. Various
theoretical attempts have been made to understand this enhancement.
In particular, an enhancement of phi meson yield may be a signature
of the formation of a quark gluon plasma in the collisions
\cite{shor}. However, this enhancement can also be explained either 
using hadronic models if one takes into account the reduced phi meson
mass in nuclear medium \cite{kosa} or via the fragmentation of
color ropes that are formed from the initial strings \cite{greiner}.

Recently, phi meson production has also been measured in central Pb+Pb 
collisions at $158A$ GeV at the CERN/SPS. The NA49 Collaboration \cite{na49} 
has identified the phi meson via the decay channel $\phi\to K^+K^-$, 
while the NA50 Collaboration \cite{na50} measured it using 
$\phi\to \mu^+\mu^-$ decay. It was found that the extracted number 
of phi meson from dimuon channel exceeds by a factor between two and four
from that extracted from the $K^+K^-$ channel. This difference 
has been attributed to the fact that not all phi mesons can be 
reconstructed from $K^+K^-$ resulting from their decays due to 
rescattering of the $K^+$ and $K^-$ in the hadronic matter. The 
suppression factor can be 40-60\% based on the RQMD model \cite{john}. 
This effect has been further studied in a schematic model by including 
the effect of changing kaon masses in the hadronic medium \cite{filip}. 

In the present paper, using a multiphase transport (AMPT) model 
we make a detailed and consistent study of the production of phi mesons 
reconstructed from the $K^+K^-$ and $\mu^+\mu^-$ decay channels.
The effects of in-medium mass modification to the phi meson yield and spectra
are also studied and are compared with the experimental findings by 
the NA49 and NA50 Collaborations at the SPS energies. We also present
the phi meson yield and spectra at the RHIC energy of $\sqrt s=130A$ GeV.

This paper is organized as follows. In Section \ref{model}, we
briefly review the AMPT model, discuss the phi meson production and
interaction cross sections in hadronic matter as well as the medium effects
on kaon and phi meson masses. The identification of phi mesons from 
both the dimuon and dikaon channels in the transport model is described
in Section \ref{phi}. In section \ref{results}, we show the
results on phi mesons from heavy ion collisions at both SPS and RHIC.
Finally, summary and conclusions are given in Section \ref{summary}.

\section{The model}\label{model}

\subsection{A Multiphase transport model}

To describe the heavy ion collision dynamics, we employ a multiphase
transport model (AMPT) that includes both the initial partonic and final 
hadronic interactions. The AMPT model \cite{ampt} is a hybrid model 
that uses as input the minijet partons from the hard processes and 
the strings from the soft processes in the HIJING model \cite{hijing}. 
The dynamical evolution of partons are then modeled by the ZPC 
\cite{zpc} parton cascade model, while the transition from the 
partonic matter to the hadronic matter is based on the Lund string 
fragmentation model \cite{lund}. The final-state hadronic
scatterings are modeled by the ART model \cite{art}.  The AMPT model 
has been quite successful in describing the measured transverse 
momenta of pions and kaons, and the rapidity distributions
of charge particles \cite{back1,acker} as well as the particle to 
antiparticle ratios \cite{back2} in heavy ion collisions at both SPS 
and RHIC. Including melting of initial strings to partons, the model 
can also account for the observed large elliptic flow \cite{lin}
and the measured two-pion correlation function \cite{hbt} at RHIC. 
In the original AMPT model, phi mesons are produced only from initial 
string fragmentation. In the present work, we extend it to also 
include phi meson production and interactions in the hadronic matter.

\subsection{phi meson production and interactions}

For phi meson production from hadron scatterings in the ART model, we 
consider both the baryon-baryon interaction channels $BB\to \phi NN$ 
and the meson-baryon channels $(\pi,\rho)B\to \phi B$, where 
$B\equiv N,\Delta,N^*$. The cross sections for these processes 
have been evaluated in the one-boson-exchange model \cite{chung} 
and are used here. We have also included phi meson production from the 
reaction $K\Lambda\to \phi N$ with the cross section taken from 
Ref. \cite{kosa} based on a kaon-exchange model. 

Phi meson can be further produced from kaon-antikaon scattering.
For a kaon-antikaon pair with invariant mass $M$, the total cross 
section for $K\bar K\to\phi$ is taken to be a Breit-Wigner form 
\begin{equation}
\sigma_{K\bar K\to \phi}(M) = \frac{3\pi}{k^2} 
\frac{(m_\phi\Gamma_\phi)^2}{(M^2-m_\phi^2)^2 + (m_\phi\Gamma_\phi) ^2} ~,
\end{equation}
where the phi meson decay width to $K\bar K$ is given by
\begin{equation}\label{gamakk}
\Gamma_{\phi\to K\bar K}(M) = \frac{g^2_{\phi K\bar K}}{4\pi} 
\frac{\left(m^2_\phi - 4m_K^2\right)^{3/2}}{6m^2_\phi} ~,
\end{equation}
with the coupling constant $g^2_{\phi K\bar K}/4\pi \approx 1.69$
determined from the empirical width of 3.7 MeV, corresponding to $83\%$
of the total width, at $M=m_\phi$.

\begin{figure}[ht]
\centerline{\epsfig{file=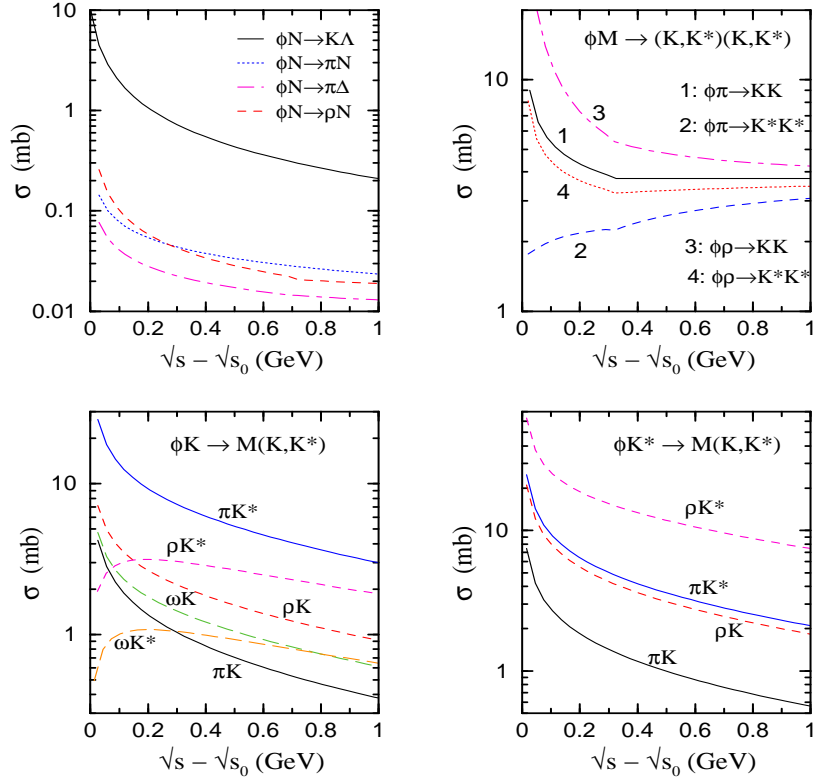,width=4.3in,height=4.3in,angle=0}}
\vspace{0.5cm}
\caption{Phi meson scattering cross sections by nucleons (upper left panel),
mesons (upper right panel), kaons (lower left panel), and $K^*$
(lower right panel).}
\label{cross}
\end{figure}

For phi meson interactions with baryons, we include the absorption 
reactions given by the inverse reactions of phi meson production from 
meson-baryon interactions given above. The corresponding 
cross sections are obtained using the detailed balance relations.
The cross section for phi meson elastic scattering with a nucleon
is taken to be 0.56 mb as extracted from the data on phi-meson 
photoproduction using the vector meson dominance model \cite{joos}. 
These cross sections are shown in the upper left panel of Fig. \ref{cross}.

Phi mesons can also be scattered by mesons. Using effective hadronic
Lagrangians, where coupling constants were determined from experimental 
partial decay rates, the total collisional width of $\phi$ due to the 
reactions $\phi\pi\to KK^*$, $\phi\rho\to KK$, $\phi K\to \phi K$ 
and $\phi\phi\to KK$ was found to be less than 35 MeV \cite{kosei}
(where $K$ denotes either a kaon or an antikaon as appropriate). 
However, recent calculations \cite{ruso} based on a Hidden 
Local Symmetry Lagrangian shows that the collisional rates of $\phi$ 
with pseudoscalar ($\pi$, $K$) and vector ($\rho$, $\omega$, 
$K^*$, $\phi$) mesons are appreciably large, especially for the $K^*$,
resulting in a  much smaller mean free path, of about 2.4 fm in a hadronic 
matter at temperature $T>170$ MeV, compared to the typical hadronic 
system size of $\sim 10-15$ fm created in heavy ion collisions.
We have included all these possible interactions, i.e., 
$\phi M \to (K,K^*)(K,K^*)$, and $\phi (K,K^*)\to M(K,K^*)$, 
where $M\equiv (\pi,\rho,\omega)$ with cross sections 
determined from the partial collisional widths given in Ref. \cite{ruso}.
Specifically, we take matrix elements $\overline{|{\cal M}|^2}$
of these reactions to be independent of the center-of-mass energy 
and extract their values from the collisional widths. The cross 
sections for these reactions are then given by  
$d\sigma/d\Omega \sim \overline{|{\cal M}|^2} k_f/(k_i s^2)$, 
where $s$ is the c.m. energy and $k_i$ and $k_f$ are, respectively, 
the initial and final c.m. momenta of the colliding hadrons.
The cross sections for the inverse reactions of the above processes 
are then obtained from the detailed balance relations.  In Fig. \ref{cross},
we show the cross sections for phi-pion and phi-rho scattering 
(upper right panel), phi-kaon scattering (lower left panel), and 
phi-$K^*$ scattering (lower right panel). The elastic cross section of 
phi meson with other mesons can be similarly obtained and is found to be 
about 2 mb. 

\subsection{medium effects}

Since the phi meson mass is only 32 MeV above the kaon-antikaon 
threshold, it is strongly coupled to the kaon and antikaon dynamics, 
i.e., to the strangeness content of the surrounding medium. 
A small change in the masses of either the phi meson or kaon 
thus would appreciably affect the decay width of phi meson.

Studies based on various approaches, such as the relativistic mean-field 
theory \cite{schaffner} and the chiral effective Lagrangian \cite{chiral},
have led to a general consensus that $K^+$ feels a rather weak repulsive 
potential while the $K^-$ feels a relatively stronger attractive potential. 
With such modifications, the kaon yield and spectra could be 
successfully explained at SIS energies (1-2$A$ GeV) \cite{ko,lilee,libro} 
and even at AGS energy ($\sim 11A$ GeV) \cite{spal}. In the present study, 
we adopt the kaon and antikaon energies from chiral effective Lagrangian, 
i.e., 
\begin{equation}\label{kmass}
\omega_{K,\bar K} = \left[m_K^2 + {\bf k}^2 - a_K\rho_S 
+ (b_K\rho)^2 \right]^{1/2} \pm b_K\rho ~,
\end{equation}
where $b_K = 3f_\pi^2/8 \approx 0.333$ GeV fm$^3$,  
$a_K \approx 0.22$ GeV$^2$ fm$^3$ for kaons, and
$a_{\bar K} \approx 0.45$ GeV$^2$ fm$^3$ for antikaons.
The $K^+$ potential, defined by $U_K=\omega_K-
\sqrt{m_K^2+{\bf k}^2}$ and given by the plus sign in Eq. (\ref{kmass}),
is then $\sim +20$ MeV at normal nuclear density and is consistent 
with that determined from the impulsive approximation using the empirical 
kaon-nucleon scattering length \cite{barnes}. The $K^-$ potential, given
by the minus sign in Eq. (\ref{kmass}), at normal nuclear density is 
about $-110$ MeV and is somewhat less than that extracted from 
the kaonic atoms \cite{fried}. The present set of parameters was found
to reproduce well the $K^+$ and $K^-$ kinetic energy spectra and the
ratio of their yields for heavy ion collisions at the SIS energy \cite{Li97}.

For the in-medium phi meson mass, the QCD sum rule studies show that
it decreases slightly in hot dense matter \cite{hatsuda,asakawa,klingl}. 
Using the result from Ref. \cite{hatsuda}, we have
\begin{equation}
\frac{m^*_\phi}{m_\phi} \approx 1 - 0.0255\frac{\rho}{\rho_0} ~,
\end{equation}
where $\rho_0$ is the normal nuclear matter density. 

The in-medium decay width of phi meson to kaon-antikaon pair is then given by 
\begin{equation}\label{gamakkm}
\Gamma^*_{\phi\to K\bar K}(M) = \frac{g^2_{\phi K\bar K}}{4\pi} 
\frac{1}{6m^{* 5}_\phi}\left[\left((m^{*2}_\phi -(m^*_K + m^*_{\bar K})^2 \right) 
\left( (m^{*2}_\phi -(m^*_K - m^*_{\bar K})^2 \right) \right]^{3/2}.
\end{equation}
Although the phi meson mass decreases in the medium, the larger reduction
of the overall kaon-antikaon mass with increasing density results in an 
increase of phi meson width $\Gamma^*_{\phi\to K\bar K}$ in the medium.
At density $\rho = 2\rho_0$, the phi meson in-medium width is about 45 MeV.

\section{Reconstruction of phi mesons}\label{phi}

Since phi meson is unstable, it can only be detected from its 
decay product of either the kaon-antikaon pair or the lepton pair.
The kaon-antikaon pair decaying from a phi meson is, however, likely
to undergo appreciable rescattering in the medium, and this would
lead to a reconstructed invariant mass situated outside the original 
phi meson peak. In fact, the momentum transfer to a kaon suffering 
one collision with a pion at $T=150$ MeV can be estimated to be about 
45 MeV which is much larger than the total phi meson width of 
$\Gamma_\phi = 4.45$ MeV. Hence phi mesons decaying in the dense
medium is difficult to be identified via reconstructed kaon-antikaon 
pairs. In contrast, dileptons have negligible final-state interactions
with the surrounding hadronic medium and therefore escape essentially 
unscathed during the entire evolution of the system. Dileptons are 
thus considered to carry useful information about hadron properties 
in hot and dense hadronic matter \cite{gqli1}, which are expected 
to be different from those in free space. 

\subsection{dileptons}

To detect a phi meson from its dimuon decay, we need the decay width
for $\phi\to \mu^+\mu^-$ \cite{dilepton}, which in the medium is
\begin{equation}\label{gamamu}
\Gamma^*_{\phi\to \mu^+\mu^-} (M) = C_{\mu^+\mu^-}
\frac{m^{* 4}_\phi}{M^3} \! \left(1 -\frac{4m^2_\mu}{M^2} \right)^{1/2}
\! \left(1 +\frac{2m^2_\mu}{M^2} \right) . 
\end{equation}
The coefficient 
$C_{\mu^+\mu^-} \equiv \alpha^2 \big/27(g^2_{\phi K\bar K}/4\pi) 
= 1.634\times 10^{-6}$ is determined from the
measured width at $M=m_\phi$. Since dimuons are emitted continuously during
the evolution of the system, the total number of dimuon is given by
\cite{dilepton}
\begin{equation}\label{muphi}
N_{\mu^+\mu^-} = \int_0^{t_f} \! dt \ N_\phi(t) 
\Gamma^*_{\phi\to \mu^+\mu^-}(M)
+ N_\phi(t_f) \frac{\Gamma_{\phi\to \mu^+\mu^-}}{\Gamma_\phi} ~,
\end{equation}
where $N_\phi(t)$ denotes the number of phi meson at time $t$. In the
above, the first term corresponds to dimuon production before the
freeze-out time of $t_f=35$ fm/c considered in our calculation, while the
second term refers to dimuon emission after freeze-out. The reconstructed
phi meson number is obtained by dividing the above expression by the 
dimuon branching ratio in free-space of  
$\Gamma_{\phi\to \mu^+\mu^-}/\Gamma_\phi = 3.7\times 10^{-4}$. Since most
previous studies \cite{john,filip} have neglected the phi meson 
annihilation and production channels via baryons and especially by 
pseudoscalar and vector mesons, the phi meson lifetime in these studies 
is thus comparable to the lifetime of the system.  Consequently, phi 
mesons from dimuon channel get significant contributions only after freeze-out.

\subsection{kaons}

The number of kaon-antikaon  pair stemming from phi meson decay can be 
similarly expressed as Eq. (\ref{muphi}) for dimuon production, i.e., 
\begin{equation}\label{kkphi}
N_{K\bar K} = \int_0^{t_f} dt \ N_\phi(t) \Gamma^*_{\phi\to K\bar K}(M)
+ N_\phi(t_f) \frac{\Gamma_{\phi\to K\bar K}}{\Gamma_\phi} ~.
\end{equation}
The phi meson abundance from kaon-antikaon decay is obtained by dividing 
Eq. (\ref{kkphi}) by the $K\bar K$ branching ratio in free-space
$\Gamma_{\phi\to K\bar K}/\Gamma_\phi$. It is evident from Eqs. (\ref{muphi})
and (\ref{kkphi}) that in absence of any medium effect on the masses and 
decay widths, i.e. for $\Gamma^*_{\phi\to K\bar K} = \Gamma_{\phi\to K\bar K}$
and $\Gamma^*_{\phi\to \mu^+\mu^-} = \Gamma_{\phi\to \mu^+\mu^-}$,
the number of reconstructed phi meson from dimuon channel is same
as that from the kaon-antikaon channel if all the kaon pairs escape 
the collision zone unscattered. On the other hand, with large in-medium 
phi meson width both the production and annihilation of phi meson from 
$\phi \leftrightarrow K\bar K$ is enhanced. If the production of phi meson 
dominates over its decay especially at the early dense stage of the 
collision, the dimuons originating from the phi meson decay will be enhanced.
On the other hand, the kaon-antikaon pairs from phi meson decays will 
be essentially undetected in the phi meson reconstruction as they are 
expected to be rescattered and experience strong medium modifications 
of their masses.

\section{results}\label{results}

We have used the extended AMPT model described in the above to
study phi meson production in heavy ion collisions at both SPS
and RHIC energies.

\subsection{SPS}

\subsubsection{time evolution}

\begin{figure}[ht]
\centerline{\epsfig{file=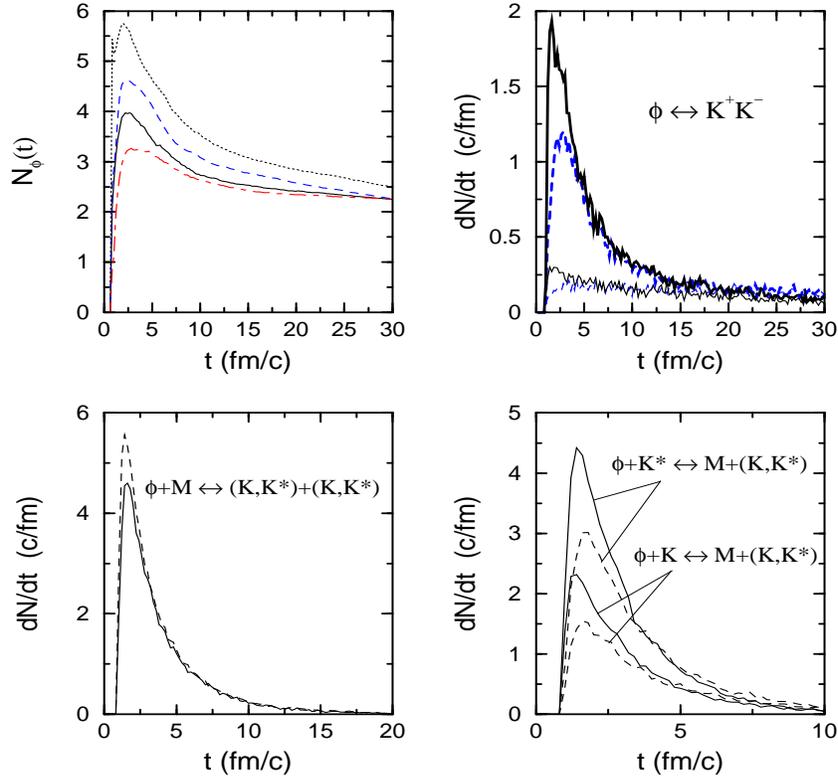,width=4.3in,height=4.3in,angle=0}}
\vspace{0.5cm}
\caption{Upper left panel: Time evolution of midrapidity ($|y|<0.5$) phi 
meson from Pb+Pb collisions at the SPS energy of $158A$ GeV at an impact 
parameter of $b\leq 3.5$ fm in the AMPT model. The results are for without
(solid curve) and with (dashed curve) in-medium mass modifications,
and with further increase of phi meson number by two in the HIJING model
(dotted curve). The phi meson yield obtained from purely hadronic rescattering
without any medium effects is shown by dash-dotted curve.
Upper right panel: The phi meson production (solid curves) 
and decay rates (dashed curves) for the process $\phi\leftrightarrow K^+K^-$. 
The thin curves are without in-medium mass modification while the thick ones 
include medium mass modification. Lower panels: $\phi$ production 
(solid curves) and absorption rates (dashed curves) for different 
channels without any mass modifications; where $M\equiv \pi,\rho,\omega$.}
\label{time}
\end{figure}

The time evolution of phi meson abundance at midrapidity for central 
Pb+Pb collisions at $158A$ GeV in the AMPT model is shown in the
upper left panel of Fig. \ref{time}. The solid curves are the results 
from calculations where both kaon and phi meson medium effects are 
neglected. The minijet partons and predominantly string decays in 
the HIJING model at the SPS energy produces about 2.6 initial number 
of phi mesons per event. Since these phi mesons are produced at 
different times as a result of finite formation time, they appear
successively in the ART hadronic transport. Subsequent multiple 
rescattering in the hadronic medium leads to a peak value of 
$N_\phi \approx 4.0$ at $t=3$ fm/c, which finally decays to 
$\sim 2.3$ at an average freeze-out time of $t_f \simeq 35$ fm/c. 
Because of small decay width of $\phi$ into $K\bar K$ pair,
most of the phi mesons decay outside the fireball (see upper right 
panel of Fig. \ref{time}). In spite of this small width, the production of 
phi meson from the inverse reaction dominates over the decay 
due to considerable abundance of kaons from the HIJING model at 
the early stage of the reaction. At times $t>15$ fm/c the process 
$\phi \leftrightarrow K\bar K$ is seen to approach chemical equilibrium. 
Note that even at late times at around the freeze-out value 
there is a substantial phi meson production from the kaon-antikaon 
channel which has been neglected in the previous study \cite{filip}.

Time evolution of the production and absorption rates of phi meson 
from different channels are shown in the lower panels of Fig. \ref{time}. 
Of all the collisional channels considered for the phi meson, 
the dominant ones are $\phi M \to (K,K^*)(K,K^*)$ and 
$\phi K^*\to M(K,K^*)$ ($M\equiv \pi, \rho,\omega$), which are of comparable 
magnitude. Although the cross section in the former reaction is smaller 
compared to that of $\phi-K^*$ (see Fig. 1), the larger abundance 
of pions relative to kaons ($K^+/\pi^+ = 0.12$ from the HIJING) 
leads to similar contribution to phi meson production in both these 
reactions. Moreover, the small abundance of phi meson results in 
identical production and annihilation rates for $\phi-M$ collision 
(lower left panel), i.e. chemical equilibrium in this reaction is 
rapidly achieved. In contrast, phi meson production overwhelms its 
destruction (lower right panel) due to larger abundances of 
the mesons $M(=\pi, \rho, \omega)$ relative to $K$ and 
especially $K^*$. In particular, the process $MK^* \to\phi K^*$ with 
its largest cross section and smaller threshold contributes $\sim 80\%$ 
to the total $\phi+K^*$ formation. This enhanced production at about 
3 fm/c is reflected in the rapid increase of the phi meson abundance 
(upper left panel). 
The production and annihilation rates for phi mesons in processes involving
$\bar K$ and $\bar K^*$, which are not shown in the figure, exhibit similar
qualitative features as for those with $K$ and $K^*$, but are about a factor 
of 1.5 smaller reflecting the $\bar K/K$ ratio of 0.70 at the SPS energy.
The phi meson collisional rate as observed here 
is considerably larger than that predicted in all previous calculations 
\cite{kosei,haglin}. As the system expands the collisional rate drops 
and these reactions reach chemical equilibration at $t=7$ fm/c. 
Finally, at a later time phi meson decay becomes more effective 
than its absorption. It may be noted that the initial dominance 
of phi meson production and its subsequent absorption as observed
in the present transport calculation is different from the findings of Ref. 
\cite{ruso}, where $\phi$ mesons embedded in a gas of ($\pi$, $K$, $\rho$, 
$\omega$, $K^*$) at  chemical equilibrium, are found to decrease 
with evolution of the system.

The chemical equilibration of phi mesons via production and annihilation 
in the hadronic matter can be better seen by switching off the initial 
phi production from string decays in the HIJING model. As shown by
the dash-dotted curve in the upper left panel of Fig. \ref{time}, 
this leads to a suppressed peak in the time evolution of phi meson 
abundance but does not change much the phi meson number at later times,
when compared to that in the case of including the initial phi mesons 
from string decays.

\subsubsection{rapidity distribution}

\begin{figure}[ht]
\centerline{\epsfig{file=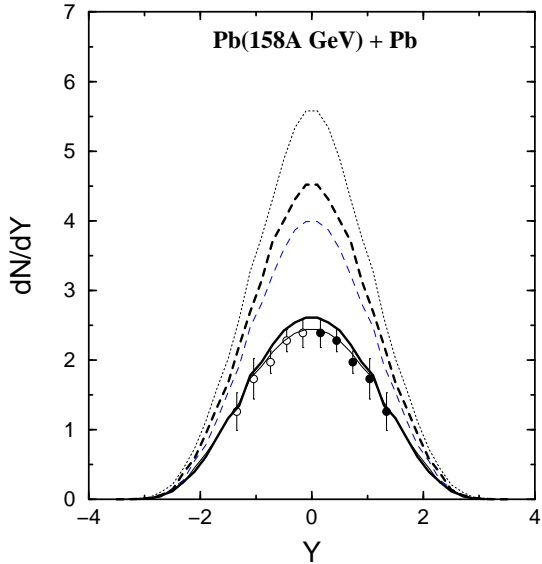,width=2.8in,height=3.0in,angle=0}}
\vspace{0.5cm}
\caption{Rapidity distribution of phi meson reconstructed from $K^+K^-$ pairs
(solid curves) and from $\mu^+\mu^-$ channel (dashed curves) for Pb+Pb 
collisions at $158A$ GeV at an impact parameter of $b\leq 3.5$ fm in 
the AMPT model. The results are for without (thin curves) and 
with (thick curves) in-medium mass modifications. The dotted curve
corresponds to phi mesons from the dimuon channel with in-medium masses 
and with the phi meson number from HIJING increased by a factor of two. 
The solid circles are the NA49 experimental data \protect\cite{na49}
from the $K^+K^-$ channel.}
\label{dndy}
\end{figure}

The rapidity distribution of phi mesons reconstructed from $K^+K^-$ and 
$\mu^+\mu^-$ channels in central Pb+Pb collision at $158A$ GeV is shown in
Fig. \ref{dndy}. In absence of any medium effects, the results for phi mesons
from the the $K^+K^-$ channel (thin solid curve) are in good agreement 
with the NA49 experimental data \cite{na49} (solid circles). With 
free-space decay width of phi meson, the rapidity distribution of all
phi mesons from the $K^+K^-$, neglecting their scattering, is identical to 
that reconstructed from the $\mu^+\mu^-$ channel (thin dashed curve).  
Due to rescattering or absorption of the kaon pairs, $\sim 30\%$ of all $\phi$
mesons are lost in the reconstruction, of which about $40\%$ of the decaying
kaons have at least one (anti)kaon that suffer elastic scattering. The maximum
depletion of phi mesons from the $K^+K^-$ channel occurs at midrapidity 
where kaon-antikaon pairs undergo appreciable scattering in the dense 
hadronic medium. Note that around $y=0$, phi meson from dimuon channel 
is about a factor 1.7 larger than from the kaon-antikaon channel, 
which is smaller than the factor of 2-4 enhancement observed in the 
NA50 data \cite{na50}.

\subsubsection{transverse mass spectrum}

\begin{figure}[ht]
\centerline{\epsfig{file=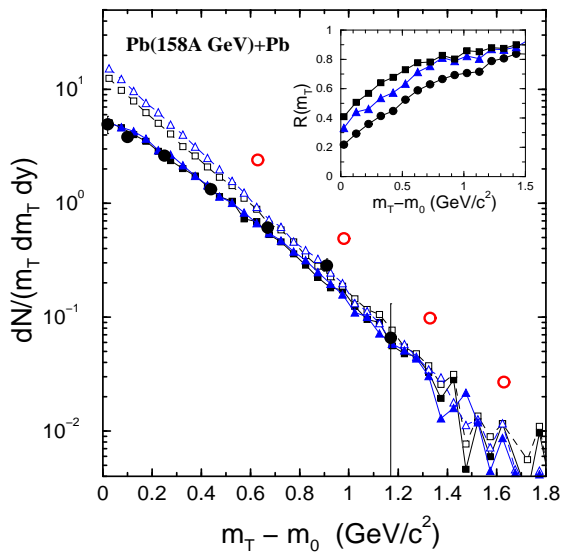,width=2.9in,height=2.9in,angle=0}}
\vspace{0.5cm}
\caption{Transverse mass spectra for midrapidity ($|y|<1$) phi mesons 
reconstructed from $K^+K^-$ pairs (solid symbols) and from $\mu^+\mu^-$ 
channel (open symbols) for Pb+Pb collisions at $158A$ GeV at an impact 
parameter of $b\leq 3.5$ fm in the AMPT model. The results are for 
without (squares) and with (triangles) in-medium mass modifications. 
The solid circles are the NA49 data \protect\cite{na49} for 
$\phi\to K^+K^-$ decay, and the open circles are the NA50 data 
\protect\cite{friese} for $\phi\to \mu^+\mu^-$.
In the inset is shown as a function of $m_T$ the ratio $R(m_T)$ for 
phi mesons decaying to kaon-antikaon pairs that are not scattered 
to those determined from the dimuon channel. The results are for 
without (squares) and with (triangles) in-medium mass modification, 
and with a further increase of phi meson number by two in the 
HIJING model (circles).}
\label{mt}
\end{figure}

In Fig. \ref{mt}, we show the transverse mass spectra at midrapidity 
for phi meson from kaon-antikaon (solid squares) and dimuon (open squares) 
channel without medium effects. It is seen that at low $m_T$ the phi 
meson from $K^+K^-$ channel is suppressed due to rescattering. Since 
the transverse momentum of a particle increases due to increasing 
number of scattering and due to pressure build-up inside the system, 
the decayed kaons at the early stages, which are predominantly scattered,
thus have low transverse momenta. The transverse mass spectra can be 
approximately fitted by $\exp(-m_T/T)$. The inverse slope parameter 
$T$ reported by NA50 \cite{na50} from the $\mu^+\mu^-$ channel for
$1.7<m_T<3.2$ GeV/c$^2$ at midrapidity is $T=227\pm 10$ MeV, which agrees
well with the AMPT model prediction of $T=228$ MeV. In contrast, the slope
parameter extracted by NA49 \cite{na49} from the $K^+K^-$ channel 
for $1<m_T<2.2$ GeV/c$^2$ is $T=305\pm 15$ MeV. The depletion of 
reconstructed $\phi$ at low $m_T$ in the AMPT model leads only to 
a slighter higher slope of $T=267$ MeV, and thus is much smaller than 
the NA49 data. The ratio of the yields at midrapidity, 
$R(m_T) = N_{K^+K^-} (\Gamma_\phi/\Gamma_{\phi\to K\bar K})\Big/
N_{\mu^+\mu^-} (\Gamma_\phi/\Gamma_{\phi\to \mu\mu})$, from Eqs. 
(\ref{kkphi}) and (\ref{muphi}) corrected by their respective branching 
ratios, is shown in the inset of Fig. \ref{mt}. The maximum suppression 
of $\sim 40\%$ at low $m_T$ in the kaon channel as observed in the 
AMPT model (squares) is larger than the suppression factor of $60\%$ 
found in the RQMD calculation \cite{john}. This can be traced back to 
enhanced phi meson production from large collisional scattering among 
the mesons leading to a peak in the $\phi$ yield at the early stage 
(see Fig. \ref{time}; upper left panel). The dimuons from these phi
meson escape the fireball freely while the $K^+K^-$ are rescattered 
in the dense hadronic medium and do not contribute to the reconstruction 
of phi mesons.

\subsubsection{In-medium effect}

Modification of phi meson and kaon masses is expected to enhance 
the production and decay of phi meson in the medium and to lead to 
a possible further increase of the suppression factor $R(m_T)$. 
The abundance of phi meson as a function of time of the colliding system 
with in-medium mass modification is shown in the upper left panel of 
Fig. \ref{time} (dashed curve). The abundance exhibits $20\%$ increase at the 
peak, and it finally merges with the free-space value at large times 
when medium effects are negligible and chemical equilibrium in all 
channels sets in. This enhancement can be mainly attributed to the
increase in the $K^+K^-\to \phi$ production rate due to larger width 
$\Gamma^*_{\phi\to K\bar K}$ in the dense medium, as shown by thick curves in 
the upper right panel of Fig. \ref{time}.

The rapidity distribution of $\phi$ from kaon pairs that have not
suffered any collision is shown in Fig. \ref{dndy} with in-medium masses 
(thick solid curve), which is found to lie within the error bars of the
NA49 data. Since free-space branching ratio is used to determine the phi meson
abundance from the kaon channel, it is evident from Eq. (\ref{kkphi}) that 
large in-medium width $\Gamma^*_{\phi\to K\bar K}$ leads to appreciable 
production and simultaneous decay of phi meson, resulting in 
$dN/dy \approx 7.1$ at $y\approx 0$ for all these kaon pairs. In contrast,
since the branching ratio for $\phi\to\mu^+\mu^-$ is largely
unaltered in the medium, the $dN/dy$ of the reconstructed phi mesons 
at $y\approx 0$ is 4.5 and thus about 1.9 times larger than 
that from the kaon channel. 

The transverse mass spectra in the kaonic channel with medium effects (solid 
triangles in Fig. \ref{mt}) reveals nearly identical slope as for the 
bare masses. On the other hand, the large number of dimuon production 
at low $m_T$ in the early stage of collisions with in-medium masses 
leads to a slightly steeper mass spectra with a slope parameter of $T= 220$ MeV. 
This is clearly seen in the inset of Fig. \ref{mt} (triangles) for the ratio 
$R(m_T)$ where the suppression at low $m_T$ is $\simeq 33\%$. 

\subsubsection{discussions}

Our results on the phi meson yield reconstructed from kaon and dimuon 
channels differs at most by about a factor of two in the AMPT model, which 
corresponds only to the lower bound to the differences found in the NA49 
and NA50 data. 
For a systematic comparison with theory it is instructive to undertake 
an experiment where both the kaon and muon pairs from phi mesons are determined
in the same $m_T$ and $y$ range. If differences as large as a factor
of four corresponding to the upper bound of the NA49 and NA50 data is 
indeed observed in the same experiment, this may then suggest that
other mechanisms, such as formation of color rope \cite{greiner}
or quark-gluon plasma \cite{shor}, are needed for phi meson production. 
In the QGP scenario, because of copious production of 
$s\bar s$ pair, its subsequent hadronization may result in a 
dramatic increase of phi meson abundance in excess of that produced purely by
hadronic rescatterings \cite{shor}. The phi meson produced at the initial
stage should contribute primarily to the dimuon channel. To mimic the effect
of enhanced phi meson production in QGP, we make an ad hoc increase 
in the phi meson number by a factor of two in the HIJING model. The time 
evolution of $N_\phi$ is illustrated in Fig. \ref{time} (dotted curve) where 
the phi meson yield, instead of exhibiting a peak, rapidly decreases to its 
equilibrium value which corroborates the findings of Ref. \cite{ruso}. 
Including also the medium effect, the rapidity distribution from 
non-scattered kaon is increased by another $\sim 8\%$, whereas the 
yield from $\mu^+\mu^-$ channel is considerably enhanced 
(dotted curve in Fig. \ref{dndy}). The resulting suppression factor 
$R(m_T)$ at low $m_T$ is $\approx 21\%$ as is evident from the inset 
of Fig. \ref{mt} (circles).

The importance of initial phi mesons to the dimuon yield is also 
seen in the scenario of neglecting phi mesons from the initial
string decays. In this case, we find that the phi number obtained 
from the $K^+K^-$ channel is essentially unaffected due to chemical 
equilibration via the hadronic scatterings, but the phi meson 
reconstructed from dimuons is, however, significantly suppressed.

\subsection{RHIC}

\begin{figure}[ht]
\centerline{\epsfig{file=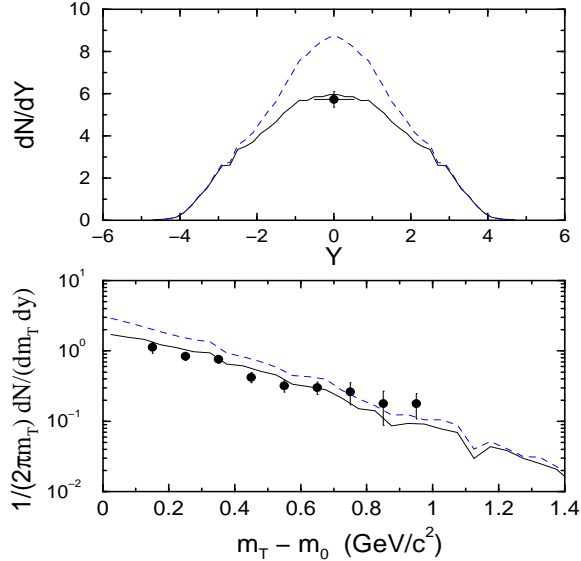,width=3.0in,height=3.0in,angle=0}}
\vspace{0.5cm}
\caption{The rapidity distribution (top panel) and the transverse mass 
spectra (bottom panel) for midrapidity ($|y|<0.5$) phi mesons reconstructed
from $K^+K^-$ pairs (solid curves) and from $\mu^+\mu^-$ channel 
(dashed curves) for Au+Au collisions at RHIC energy of $\sqrt s =130A$ 
GeV at an impact parameter of $b\leq 5.3$ fm in the AMPT model. 
The solid circles are the STAR experimental data \protect\cite{star}
for $0-11\%$ central collisions for $\phi$ reconstructed from $K^+K^-$ decay.} 
\label{dndymt}
\end{figure}

In the AMPT model, heavy ion collisions at RHIC differ from that
at SPS mainly in the increasing effect of minijet gluons. However,
the initial heavy ion collision dynamics at RHIC remains dominated by  
soft string fragmentation processes as at SPS. 
In Fig. \ref{dndymt} (top panel), we present 
results from the AMPT model for the rapidity distribution of phi mesons
reconstructed from $K^+K^-$ and $\mu^+\mu^-$ pairs for central Au+Au 
collisions at the RHIC energy of $\sqrt s=130A$ GeV. Of the total 
phi meson yield of 37 per event, about $24\%$ of these are lost in the kaonic 
channel by hadronic rescattering and absorption. Note that in spite of 
enhanced phi meson production in the early stages of the collision
as compared to the SPS energy, its abundance at $y\approx 0$ is only about a
factor of 1.5 large in the dimuon channel. This may be attributed to 
fewer kaons lost by rescattering at RHIC as they can escape rapidly out of the 
collision zone unperturbed due to of their large energies. It is seen that
phi meson multiplicities reconstructed from $K^+K^-$ pairs at midrapidity 
in the AMPT model is consistent with the STAR data \cite{star}.

The transverse mass spectra of $\phi$ meson from the two channels are 
shown in the bottom panel of Fig. \ref{dndymt}, and compared with the 
STAR data \cite{star} for the $K^+K^-$ channel. Compared to a slope 
parameter of $T=379\pm 50$ MeV in the data, the AMPT model predicts 
a smaller value of $T=335$ MeV in the kaonic channel in the range 
$0<m_T-m_\phi<1$ GeV. The slope parameter for phi mesons determined 
from the $\mu^+\mu^-$ channel is $T=297$ MeV resulting in a suppression 
factor at low $m_T$ of $R(m_T) = 58\%$.  Note that at both SPS and 
RHIC energy, the slope parameter of phi mesons from $K^+K^-$ decay 
is smaller compared to the experimental data. We have increased the phi 
meson elastic cross sections with baryons and mesons to 8.3 mb, which 
corresponds to the upper bound estimated from phi meson photoproduction 
data \cite{chung}, 
and find that this increases only slightly the slope parameter. On the 
other hand, an enhanced flow is expected to be generated by converting
the strings produced from soft processes into interacting partons 
in the high energy density created in ultra-relativistic heavy ion 
collisions \cite{lin}. Since in the default HIJING model the strings 
are basically noninteracting, the latter modification was found to 
reproduce the large elliptic flow observed in Au+Au collisions at 
$\sqrt s = 130A$ GeV \cite{acker}. Results based on this new AMPT
model will be reported in the future.

\section{summary and conclusions}\label{summary}

In summary, we have investigated in a multiphase transport (AMPT) model the
production of phi mesons reconstructed from $K^+K^-$ and $\mu^+\mu^-$ decays.
Considering all possible collisions of $\phi$ with the hadronic medium,
we find the phi meson yield from dimuon channel is about a factor of 1.7
higher than the kaon-antikaon channel for central Pb+Pb collisions at 
the SPS energy. The kaons originating from phi meson decay in the dense 
hadronic medium are predominantly scattered and absorbed and thus do not 
contribute to the phi meson yield, whereas all the $\mu^+\mu^-$ pair 
freely escape. Inclusion of in-medium modification of masses of kaons 
and phi mesons enhances the phi meson width and results in a factor of 
two increase of the phi meson yield from the dimuon channel
compared to the kaon-antikaon channel. However, strikingly different 
results were found by two experiments: The NA50 \cite{na50} measured 
$\phi\to\mu^+\mu^-$ and found its yield a factor of 2-4 larger with a 
smaller slope parameter compared to $\phi\to K^+K^-$ measurements 
by NA49 \cite{na49} for central Pb+Pb collisions at $158A$ GeV. 
Such a large enhancement of four at low transverse mass from dimuon 
over the kaon channel could only be explained in the AMPT model by 
an ad hoc increase of phi meson number by about three from the initial 
stage. This may be suggestive of other mechanisms, such as the 
formation of color ropes or quark-gluon plasma in these collisions.  
To pin down these differences an experiment measuring phi meson from 
both dimuon and kaon decays in the same kinematic range is therefore 
called for. For heavy ion collisions at RHIC energy of $\sqrt s= 130A$ GeV, 
the AMPT model gives the yield and slope parameter of phi mesons
from the $K^+K^-$ channel that are in reasonable agreement with 
the STAR data. Similar to the results for heavy ion collisions
at SPS, the phi meson yield from the dimuon channel is about
1.5 times larger than that from the $K^+K^-$ channel in the absence
of additional phi meson production due to the formation of the
initial partonic stage. This number is expected to increase
significantly if we include such enhanced production of initial phi 
mesons. Indeed at RHIC, large energy densities are expected to be reached, 
and a partonic matter with a longer lifetime and occupying a larger volume 
could be formed. Compared to SPS energy, larger $s\bar s$ production in the 
partonic stage and its subsequent hadronization would therefore result in an 
enhanced phi meson abundance, especially at the early stages. These phi mesons
should contribute dominantly in the dimuon channel while most of them will
escape detection in the kaonic channel. It is thus of great interest to have 
experimental data on phi mesons from the dimuon measurement at the RHIC 
energy. 

\section{acknowledgment}

This paper is based on work supported by the National Science Foundation 
under Grant Nos. PHY-9870038 and PHY-0098805, the Welch Foundation under 
Grant No. A-1358, and the Texas Advanced Research Program under Grant No. 
FY99-010366-0081.


\end{document}